\documentclass[sigconf, nonacm]{acmart}

\renewcommand\footnotetextcopyrightpermission[1]{}

\usepackage{amsmath, amsthm, amssymb, amsfonts} \usepackage{algpseudocode} \usepackage{algorithm} \usepackage{graphicx} \usepackage{multirow} \usepackage{enumitem}
\usepackage{setspace}

\AtBeginDocument{%
	\providecommand\BibTeX{{%
			\normalfont B\kern-0.5em{\scshape i\kern-0.25em b}\kern-0.8em\TeX}}}

\begin{document}
\title{HSR: Hyperbolic Social Recommender}

\author{Anchen Li}
\affiliation{%
	\institution{College of Computer Science and Technology, Jilin University, China}}
\email{liac20@mails.jlu.edu.cn}

\author{Bo Yang}
\authornote{Corresponding author.}
\affiliation{%
	\institution{College of Computer Science and Technology, Jilin University, China}}
\email{ybo@jlu.edu.cn}


\begin{abstract}
	
With the prevalence of online social media, users' social connections have been widely studied and utilized to enhance the performance of recommender systems. In this paper, we explore the use of hyperbolic geometry for social recommendation. We present \textbf{H}yperbolic \textbf{S}ocial \textbf{R}ecommender (HSR), a novel social recommendation framework that utilizes hyperbolic geometry to boost the performance. With the help of hyperbolic spaces, HSR can learn high-quality user and item representations for better modeling user-item interaction and user-user social relations. Via a series of extensive experiments, we show that our proposed HSR outperforms its Euclidean counterpart and state-of-the-art social recommenders in click-through rate prediction and top-$K$ recommendation, demonstrating the effectiveness of social recommendation in the hyperbolic space.

\end{abstract}

\maketitle
\section{Introduction}
In the era of information explosion, recommender systems have been playing an indispensable role in meeting user preferences by recommending relevant items. Collaborative Filtering (CF) is one of the dominant techniques used in recommender systems \cite{CF1,CF2,CF3,CF4}. Traditional CF-based methods mainly rely on the history of user-item interaction to generate recommendations. However, they are often impeded by data sparsity and cold start issues in real recommendation scenarios.

\par
With the emergence of online social media, many E-commerce sites have become popular social platforms in which users can not only select items they love but also follow other users. According to the social influence theory \cite{social_theory1,social_theory2,social_theory3}, users’ preferences are similar to or influenced by their social neighbors. Therefore, researchers propose using social network as another information stream to mitigate the lack of user-item interaction and enhance recommendation performance, also known as the social recommendation \cite{SoRec}. To tackle the social recommendation problem, a diverse plethora of social-aware models have been proposed, which utilizes different techniques to integrate social relations into recommendation, such as matrix factorization \cite{SocialReg,TrustSVD}, multi-layer perceptron \cite{DeepSoR} and graph neural networks \cite{Diffnet,HOSR}. 

Notably, all of these social recommenders operate in Euclidean spaces. In a real-world scenario, user-item interaction and user-user social relations exhibit power-law structures. Also, social networks present a hierarchical structure \cite{social_h3,social_h2,social_h1}. Recent research shows that hyperbolic geometry enables embeddings with much smaller distortion when embedding data with the power-law distribution and hierarchical structure \cite{Poincare,HGCN}. This motivates us to consider whether we can utilize hyperbolic geometry for boosting performance of social recommendation.

\par
\textit{\textbf{Our approach.}} In this work, we propose a novel social recommendation model, namely, \textbf{H}yperbolic \textbf{S}ocial \textbf{R}ecommender (HSR). Specifically, HSR learn user and item representations in the hyperbolic space – or more precisely in a Poincar$\acute{\text{e}}$ ball. The key component of our framework is that we design a hyperbolic aggregator on the users' social neighbor sets to take full advantage of the social information. With the help of hyperbolic geometry, HSR can better model user-item interaction and user-user social relations to enhance the performance in social recommendation.

\par
\textit{\textbf{Our contributions.}} In summary, our main contributions in this paper are listed as follows:
\begin{itemize}
	\item We propose a novel framework HSR, which utilizes hyperbolic geometry for social recommendation. To the best of our knowledge, this is the first study to make use of hyperbolic space for social recommendation task.
	\item Experimental results show that our proposed HSR not only outperforms its Euclidean counterpart but also boosts the performance over the state-of-the-art social recommenders in click-through rate prediction and top-$K$ recommendation, demonstrating the effectiveness of social recommendation in hyperbolic geometry.
\end{itemize}

The remainder of this paper is organized as follows. Section~\ref{sec:background} discusses the relevant background that forms the basis of our work. In Section~\ref{sec:methodology}, we introduce the proposed method HSR. In Section~\ref{sec:experiment}, we conduct experiments on four real-world datasets and present the experimental results. In Section~\ref{sec:related}, we review work related to our methods, followed by a conclusion in Section~\ref{sec:conclusion}.

\section{Background}
\label{sec:background}
In this section, we review the background of hyperbolic geometry and gyrovector space, which forms the basis of our method.

\subsection{Hyperbolic Geometry}
The hyperbolic space is uniquely defined as a complete and simply connected Riemannian manifold with constant negative curvature \cite{hp1}. A key property of hyperbolic spaces is that they expand faster than Euclidean spaces. To describe the hyperbolic space, there are multiple commonly used models of hyperbolic geometry, such as the Poincar$\acute{\text{e}}$ model, hyperboloid model, and Klein model \cite{hp2}. These models are all connected and can be converted into each other. In this paper, we work with the Poincar$\acute{\text{e}}$ ball model because it is well-suited for gradient-based optimization \cite{Poincare}. 

\textit{Poincar$\acute{\text{e}}$ ball model.}
Let $\mathbb{D}^n = \left\{  \textbf{x} \in \mathbb{R}^n: \left \| \textbf{x} \right \| < 1  \right\}$ be the $open$ $n$-dimensional unit ball, where $\left \| \cdot \right \|$ denotes the Euclidean norm. The Poincar$\acute{\text{e}}$ ball model is the Riemannian manifold $(\mathbb{D}^n, g^{\mathbb{D}})$, which is defined by the manifold $\mathbb{D}^n$ equipped with the Riemannian metric tensor $g_{\textbf{x}}^{\mathbb{D}} = \lambda_\textbf{x}^2g^{\mathbb{E}}$, where $\lambda_\textbf{x} = \frac{2}{1-\left \| \textbf{x} \right \|^2}$; $\textbf{x} \in \mathbb{D}^n$; and $g^{\mathbb{E}} = \textbf{I}$ denotes the Euclidean metric tensor.

\subsection{Gyrovector Spaces}
The framework of gyrovector spaces provides vector operations for hyperbolic geometry \cite{HNN}. We will make extensive use of these gyrovector operations to design our model. Specifically, these operations in gyrovector spaces are defined in an open $n$-dimensional ball $\mathbb{D}^n_c = \left\{  \textbf{x} \in \mathbb{R}^n: c \left \| \textbf{x} \right \|^2 < 1  \right\}$ of radius $\frac{1}{\sqrt{c}}(c \geq 0 )$. Some widely used vector operations of gyrovector spaces are defined as follows:
\begin{itemize}[leftmargin= 12 pt]
	\item \textit{M{\"o}bius addition}: For $\textbf{x}, \textbf{y} \in \mathbb{D}^n_c$, the M{\"o}bius addition of $\textbf{x}$ and $\textbf{y}$ is defined as follows:
	\begin{align}
	\textbf{x} \oplus_c \textbf{y} = \frac{(1+2c\left \langle \textbf{x},\textbf{y} \right \rangle + c\left \| \textbf{y} \right \|^2)\textbf{x} + (1-c\left \| \textbf{x} \right \|^2)\textbf{y}}{1+2c\left \langle \textbf{x},\textbf{y} \right \rangle + c^2\left \| \textbf{x} \right \|^2 \left \| \textbf{y} \right \|^2}.
	\end{align}
	In general, this operation is not commutative nor associative. 
	\item \textit{M{\"o}bius scalar multiplication}: For $c > 0$, the M{\"o}bius scalar multiplication of $\textbf{x} \in \mathbb{D}^n_c \backslash \left \{ \textbf{0} \right \}$ by $r \in \mathbb{R}$ is defined as follows:
	\begin{align}
	r \otimes_c \textbf{x} = \frac{1}{\sqrt{c}}\tanh\left ( r \tanh^{-1} \left ( \sqrt{c} \left \| \textbf{x} \right \| \right ) \right ) \frac{\textbf{x}}{\left \| \textbf{x} \right \|},
	\end{align}
	and $r \otimes_c \textbf{0} = \textbf{0}$. This operation satisfies associativity: 
	\item \textit{M{\"o}bius matrix-vector multiplication}: For $\textbf{M} \in \mathbb{R}^{n' \times n}$ and $\textbf{x} \in \mathbb{D}^n_c$, if $\textbf{Mx} \neq \textbf{0}$, the M{\"o}bius matrix-vector multiplication of $\textbf{M}$ and $\textbf{x}$ is defined as follows:
	\begin{align}
	\textbf{M} \otimes_c \textbf{x} = \frac{1}{\sqrt{c}}\tanh\left ( \frac{\left \| \textbf{Mx} \right \|}{\left \| \textbf{x} \right \|} \tanh^{-1} \left ( \sqrt{c} \left \| \textbf{x} \right \| \right ) \right ) \frac{\textbf{Mx}}{\left \| \textbf{Mx} \right \|}.
	\end{align}
	This operation satisfies associativity.
	\item \textit{M{\"o}bius exponential map and logarithmic map}: For $\textbf{x} \in \mathbb{D}^n_c$, it has a tangent space $T_\textbf{x}\mathbb{D}^n_c$ which is a local first-order approximation of the manifold $\mathbb{D}^n_c$ around $\textbf{x}$. The logarithmic map and the exponential map can move the representation between the two manifolds in a correct manner. For any $\textbf{x} \in \mathbb{D}^n_c$, given $\textbf{v} \neq \textbf{0}$ and $\textbf{y} \neq \textbf{x}$, the M{\"o}bius exponential map $\exp_{\textbf{x}}^c: T_\textbf{x}\mathbb{D}^n_c \rightarrow \mathbb{D}^n_c$ and logarithmic map $\log_{\textbf{x}}^c: \mathbb{D}^n_c \rightarrow T_\textbf{x}\mathbb{D}^n_c$ are defined as follows:
	\begin{align} 
	\exp_{\textbf{x}}^c(\textbf{v}) = \textbf{x} \oplus_c \left(\tanh\left(\sqrt{c}\frac{\lambda_\textbf{x}^c\left \| \textbf{v} \right \|}{2}\right)\frac{\textbf{v}}{\sqrt{c}\left \| \textbf{v} \right \|}\right),
	\end{align}
	\begin{align}
	\log_{\textbf{x}}^c(\textbf{y}) = \frac{2}{\sqrt{c}\lambda_\textbf{x}^c} \tanh^{-1}(\sqrt{c}\left \| -\textbf{x} \oplus_c \textbf{y} \right \|) \frac{-\textbf{x} \oplus_c \textbf{y}}{\left \| -\textbf{x} \oplus_c \textbf{y} \right \|},
	\end{align}
	where $\lambda_\textbf{x}^c  = \frac{2}{1-c\left \| \textbf{x} \right \|^2}$ is the conformal factor of $(\mathbb{D}^n_c, g^c)$, where $g^c$ is the generalized hyperbolic metric tensor. 
	\item \textit{Distance}: For $\textbf{x}, \textbf{y}\in \mathbb{D}^n_c$, the generalized distance between them in Gyrovector spaces are defined as follows:
	\begin{align}
	d_{c}(\textbf{x}, \textbf{y}) = \frac{2}{\sqrt{c}}\tanh^{-1}(\sqrt{c}\left \| -\textbf{x} \oplus_c \textbf{y} \right \|).
	\end{align}
\end{itemize}

We will make use of these M{\"o}bius gyrovector space operations to design our recommendation framework.

\section{METHODOLOGY}
\label{sec:methodology}
In this section, we first introduce the notations and formulate the problems. We then elaborate our proposed HSR method. Based on that, we introduce two strategies to further improve our model. Finally, we discuss the learning algorithm of our model.

\subsection{Problem Definition}
In a typical recommendation scenario, we suppose there are $M$ users $\mathcal{U}$ = $\left\{ u_{1},u_{2},...,u_{M} \right\}$ and $N$ items $\mathcal{V}$ = $\left\{ v_{1},v_{2},...,v_{N} \right\}$. We define \textbf{Y} $\in \mathbb{R}^{M \times N}$ as the user-item historical interaction matrix whose element $y_{ai} = 1$ if user $a$ is interested in item $i$ and zero otherwise. In addition to the interaction matrix \textbf{Y}, we also have a social network \textbf{S} $\in \mathbb{R}^{N \times N}$ among users $\mathcal{U}$, where its element $s_{ab} = 1$ if $u_{a}$ trust $u_{b}$ and zero otherwise. 

\par
The goal of our method HSR is to utilize the historical interaction and the social information to predict user’s personalized interests in items. Specifically, given the user-item interaction matrix \textbf{Y} as well as user social network \textbf{S}, HSR aims to learn a prediction function $\hat{y}_{ai} = \mathcal{F}(u_a, v_i | \Theta, \textbf{Y}, \textbf{S})$, where $\hat{y}_{ai}$ is the preference probability from user $u_a$ to item $v_i$ which she has never engaged before, and $\Theta$ is the model parameters of function $\mathcal{F}$.

\subsection{Model Formulation}
We now present the our method HSR. There are three components in the model: the embedding layer, the aggregation layer, and the prediction layer. Details of each part are described in the following.

\subsubsection{Embedding Layer}
Such layer takes a user and an item as an input, and encodes them with dense low-dimensional embedding vectors. Specifically, given one hot representations of target user $u_a$ and target item $v_i$, the embedding layer outputs their embeddings $\textbf{u}_a$ and $\textbf{v}_i$, respectively. We will learn user and item embedding vectors in the hyperbolic space $\mathbb{D}^d_c$. 

\subsubsection{Aggregation Layer}
Due to the social influence theories \cite{social_theory1,social_theory2,social_theory3}, user’s preference will be indirectly influenced by her social friends. We should utilize social information for better user embedding modeling. Specially, we devise a social aggregator on the users' trusted neighbors to refine users' embeddings in the hyperbolic space, formulating the aggregation process with two major operations: \textit{neighbor aggregation} and \textit{feature update}.

The neighborhood aggregation stage learns the representation of a neighborhood by transforming and aggregating the feature information from the neighborhood. The center-neighbor combination stage learns the representation of the central node by combining the representation of the neighborhood with the features of the central node.

\paragraph{Neighbor aggregation}Given a user, neighbor aggregation first aggregates the given user's neighbor representations into a single embeddings, and then combines the representation of the neighborhood with the feature of the given user. We can directly utilize M{\"o}bius addition to achieve the neighbor aggregation. Denote $\mathcal{N}(a)$ as user $u_a$'s social neighbor set, the neighbor aggregation for user $u_a$ is defined as follows:
\begin{equation} 
\begin{split} 
\textbf{u}_{\mathcal{N}(a)} &=  \sum\nolimits_{b \in \mathcal{N}(a)}^{\oplus_c} \textbf{u}_b \\
\textbf{u}^{AGG}_a &= \textbf{u}_a \oplus_c \left ( \gamma \otimes_c \textbf{u}_{\mathcal{N}(a)} \right )
\label{eq:limit_agg}
\end{split}
\end{equation}
where $\sum\nolimits^{\oplus_c}$ is the accumulation of M{\"o}bius addition, and $\gamma$ is a coefficient which controls the social influence.

\paragraph{Feature update}
In this stage, we further update the aggregated representations to obtain sufficient representation power as:
\begin{align}
\textbf{u}_a^* = \sigma\left ( \textbf{M} \otimes_c \textbf{u}^{AGG}_a \right ) 
\end{align}
where $\textbf{M}\in \mathbb{R}^{d \times d}$ is the trainable matrix, and $\sigma$ is the nonlinear activation function defined as LeakyReLU \cite{LeakyReLU}.

\paragraph{High-order aggregation}
\par
Through a single aggregation layer, user representation is dependent on itself as well as the direct social neighbors. We can further stack more layers to obtain high-order information from multi-hop neighbors of users. More formally, in the $\ell$-th layer, for user $u_a$, her representation is defined as:
\begin{align} 
\textbf{u}_a^\ell = \sigma\left ( \textbf{M}^{\ell} \otimes_c\left (\textbf{u}_a^{\ell-1} \oplus_c \left ( \gamma \otimes_c \sum\nolimits_{b \in \mathcal{N}(a)}^{\oplus_c} \textbf{u}_b^{\ell-1} \right )\right ) \right ) 
\end{align}
where $\textbf{M}^{\ell}$ is the trainable matrix in the $\ell$-th layer and $\textbf{u}_a^1 = \textbf{u}_a^*$.

\subsubsection{Prediction Layer}
After the $L$ aggregation layer, for target user $u_a$, we obtain her final representation $\textbf{u}_a^L$. We then feed user representation $\textbf{u}_a^L$ and target item representation $\textbf{v}_i$ into a function $p$ for predicting the probability of $u_a$ engaging $v_i$: $\hat{y}_{ai} = p(\textbf{u}_{a}^L, \textbf{v}_{i})$. Here we implement the prediction function $f$ as the Fermi-Dirac decoder \cite{hp1,Poincare}, a generalization of sigmoid function, to compute probability scores between $u_a$ and $v_i$, as follows:
\begin{align}
\hat{y}_{ai} = \frac{1}{e^{(d_c(\textbf{u}_a^L, \textbf{v}_i) -r)/t}+1}
\label{eq:predict}
\end{align}
where $r$ and $t$ are hyper-parameters. 

\subsection{Model Improvement}
\par
In this subsection, we further improve our method from two aspects. 

\subsubsection{Acceleration strategy}
Since M{\"o}bius addition operation is not commutative nor associative \cite{HNN}, we have to calculate the accumulation of M{\"o}bius addition by order in Equation ~(\ref{eq:limit_agg}), as follows: 
\begin{equation}
\begin{split}  
\textbf{u}^{AGG}_a 
&= \textbf{u}_a \oplus_c \left ( \gamma \otimes_c \sum\nolimits_{b \in \mathcal{N}(a)}^{\oplus_c} \textbf{u}_b \right ) \\
&= \textbf{u}_a \oplus_c \Big ( \gamma \otimes_c \big( \left((\textbf{u}_{b_1} \oplus_c \textbf{u}_{b_2}) \oplus_c \textbf{u}_{b_3}\right) \oplus_c \cdots \big)\Big) 
\label{eq:limit_agg2}
\end{split}
\end{equation}
As is known to all, there exist some active users who have many social behaviors in social network, so the calculation in Equation ~(\ref{eq:limit_agg2}) is seriously time-consuming, which will affect the efficiency of our method HSR. Therefore, it is necessary to devise a new way to calculate the aggregation. 

\par
\begin{figure}[h!]
	\centering
	\includegraphics[width=0.94\linewidth]{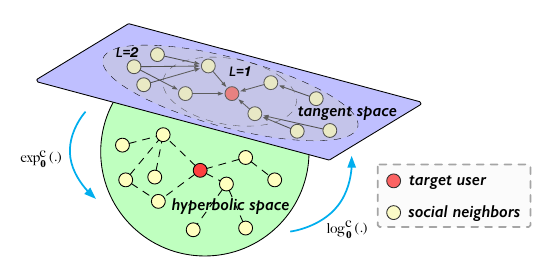}
	\caption{Illustration of the acceleration strategy of second-order embedding propagation for the target user.}
	\label{fig:map}
\end{figure}

 
We resort to M{\"o}bius logarithmic map and exponential map, as illustrated in Figure~\ref{fig:map}. Specifically, we first utilizes the logarithmic map to project user representations into a tangent space, then perform the accumulation operation to aggregate the user representations in the tangent space, and finally project aggregated representations back to the hyperbolic space with the exponential map. The process is formulated as:
\begin{align}
\textbf{u}^{AGG}_{a} = \exp_{\textbf{0}}^c \Big ( \log_{\textbf{0}}^c(\textbf{u}_a) + \gamma \cdot \sum_{b \in \mathcal{N}(a)} \log_{\textbf{0}}^c(\textbf{u}_b) \Big )
\label{eq:eff_agg}
\end{align}
Different from Equation ~(\ref{eq:limit_agg}), we can calculate the results in a parallel way in Equation ~(\ref{eq:eff_agg}) because the accumulation operation in the tangent space is commutative and associative, which enables our model more efficient. 

\subsubsection{Attention mechanism}
For a user, social influence strength of her friends should be different and dynamic. Therefore, we introduce an attention mechanism, which assigns non-uniform weights to users’ social neighbors. We donate $\pi_{ab}$ as the social influence strength from user $u_a$'s social neighbor $u_b$ to $u_a$, which can be compute as follows:
\begin{align} 
\pi_{ab} \!=\! \left ( \log_{\textbf{0}}^c(\textbf{u}_a) \!\odot\! \log_{\textbf{0}}^c(\textbf{u}_b)\right )^\top\!\tanh\left ( \textbf{w}^\top\!\left [\log_{\textbf{0}}^c(\textbf{u}_{b}), \log_{\textbf{0}}^c(\textbf{v}_i) \right ] \right )
\label{eq:att}
\end{align}
where $\left [  \cdot, \cdot\right ]$ and $\odot$ mean concatenation operation and element-wise product between two vectors in the tangent space. $\textbf{w}\in \mathbb{R}^{2d \times d}$ is the trainable weighted matrix of the attention mechanism. We also employ $\tanh$ as the nonlinear activation function. 
\par
We consider target user, target item and target user's social neighbor to design the attention mechanism. The first term calculates the compatibility between user $u_a$ and her social neighbor $u_{b}$, and the second term computes the opinions of the neighbor $u_{b}$ on the target item $v_i$. Here we simply employ inner product on the two terms, one can design a more sophisticated attention mechanism.

\begin{figure}[h!]
	\centering
	\includegraphics[width=0.9\linewidth]{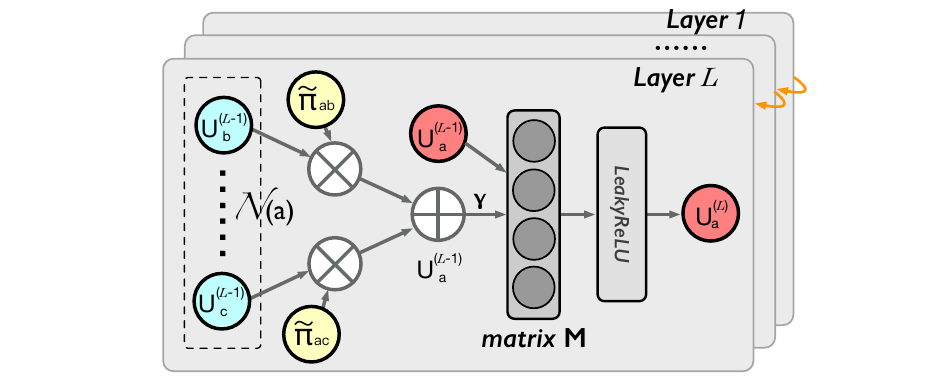}
	\caption{Illustration of the aggregation operation.}
	\label{fig:agg}
\end{figure}

\par By integrating Equation ~(\ref{eq:eff_agg}) and ~(\ref{eq:att}), the final aggregation rule (as illustrated in Figure~\ref{fig:agg}) for each user $u_a$ is calculated as follows:
\begin{align}
\textbf{u}_a^\ell \!=\! \sigma\!\Big ( \textbf{M}^\ell \!\otimes_c\!\exp_{\textbf{0}}^c \!\Big ( \log_{\textbf{0}}^c(\textbf{u}_a^{\ell-1}) + \gamma \cdot\!\!\! \sum_{b \in \mathcal{N}(a)} \!\tilde{\pi}_{ab} \!\cdot \log_{\textbf{0}}^c(\textbf{u}_b^{\ell-1}) \Big )\! \Big ) 
\end{align}
where $\tilde{\pi}_{ab}$ is the normalized attention weight which can be calculated by softmax function with temperature parameter $\tau$:
\begin{align}
\tilde{\pi}_{ab} = \frac{\exp(\pi_{ab}/\tau)}{\sum_{_{b'} \in \mathcal{N}(a)}\exp(\pi_{ab'}/\tau) }
\end{align}

\subsection{Model Optimization}

\subsubsection{Objective Function}
The complete objective function consists of two parts: recommendation loss and social relation loss.

\paragraph{Recommendation Loss}
For the recommendation task, the loss function of recommendation is defined as:
\begin{align} 
\mathcal{L}_{r} = - \sum_{(u, i, j)\in\mathcal{D}_r} \left ( y_{ui}\log(\hat{y}_{ui}) + (1-y_{uj})\log(1-\hat{y}_{uj}) \right )
\end{align}
where $\mathcal{D}_r$ is the set of training triplets. We donate $\mathcal{V}_u$ is the item set which user $u$ has interacted with, and $\mathcal{D}_r$ can be defined as:
\begin{align} 
\mathcal{D}_r = \left \{ (u, i, j) \ | \  u\in\mathcal{U} \land i\in\mathcal{V}_u \land j\in\mathcal{V}\backslash\mathcal{V}_u \right \}
\label{eq:Dr}
\end{align}

\paragraph{Social Relation Loss}
\par
To model user social relations in social network, we devise the social relation loss. We first define the score (similarity) function for a user pair $(u_a, u_b)$ according to their distance in the hyperbolic space, as follows:
\begin{align} 
score(u_a, u_b) = - \ d_c(\textbf{u}_a, \textbf{u}_b)
\label{eq:social_score}
\end{align}

\par
We want the observed trusted user pairs to have higher scores than unobserved user pairs in social network. We utilize the Bayesian personalized ranking (BPR) method to define $\mathcal{L}_{s}$, as follows:
\begin{align} 
\mathcal{L}_{s} = - \sum_{(u, p, q) \in \mathcal{D}_s} \log\sigma\left(score(u, p) - score(u, q)\right)
\label{eq:social_loss}
\end{align}
where $\sigma$ is the sigmoid function $\sigma\left(x\right) = 1/\left( 1 + e^{-x} \right)$, and $\mathcal{D}_s$ is defined as follow:
\begin{align} 
\mathcal{D}_s = \left \{ (u, p, q) \ | \  u\in\mathcal{U} \land p\in\mathcal{N}_{(u)} \land q\in\mathcal{U}\backslash\mathcal{N}_{(u)} \right \}
\label{eq:Ds}
\end{align}

\par
\paragraph{Multi-Task Learning}
We integrate the recommendation loss $\mathcal{L}_{r}$ and the social relation loss $\mathcal{L}_{s}$ into an end-to-end fashion through a multi-task learning framework, and utilize $\lambda$ to balance two terms. The complete loss function of HSR is defined as follows:
\begin{equation} 
\label{equ:loss_high}
\underset{\Theta}{\min} \mathcal{L} = \mathcal{L}_{r} + \lambda \mathcal{L}_{s}
\end{equation}
where $\Theta $ is the total parameter space, including user embeddings $\left \{ \textbf{u}_i \right \}_{i=1}^{\!|\mathcal{U}|}$, item embeddings $\left \{ \textbf{v}_i \right \}_{i=1}^{\!|\mathcal{V}|}$, and weight parameters of the networks $\left \{ \textbf{M}^i, \textbf{w}^i \right \}_{i=1}^{L}$.

\subsubsection{Gradient Conversion}
Since the Poincar$\acute{\text{e}}$ Ball has a Riemannian manifold structure, we utilize Riemannian stochastic gradient descent (RSGD) to optimize our model \cite{RSGD}. As similar to \cite{Poincare}, the parameter updates are of the following form:
\begin{align} 
\theta_{t+1} = \mathfrak{R}_{\theta_t}(-\eta_t\nabla_R\mathcal{L}(\theta_t)),
\end{align}
where $\mathfrak{R}_{\theta_t}$ denotes a retraction onto $\mathbb{D}$ at $\theta$ and $\eta_t$ denotes the learning rate at time $t$. The Riemannian gradient $\nabla_R$ can be computed by rescaling the Euclidean gradient $\nabla_E$ with the inverse of the Poincar$\acute{\text{e}}$ ball metric tensor as $\nabla_R = \frac{ ( 1-\left \| \theta_t \right \|^2 )^2}{4} \nabla_E$.

\subsubsection{Learning Algorithm}
The training process of the HSR is summarized in Algorithm~\ref{alg:1}. 

\begin{algorithm}[h!] 
	\caption{Training algorithm of HSR}
	\begin{algorithmic}
		\Require Interaction matrix $\textbf{Y}$; social network $\textbf{S}$
		\Ensure Prediction function $\mathcal{F}(u, v | \Theta, \textbf{Y}, \textbf{S})$
	\end{algorithmic}
	\begin{algorithmic}[1]
		\State Initialize all parameters in $\Theta$
		\Repeat
		\State Draw a mini-batch of $(u, i, j)$ from $\mathcal{D}_r$
		\State Draw a mini-batch of $(u, p, q)$ from $\mathcal{D}_s$
		\State Calculate $\mathcal{L}_r$ according to Equation ~(\ref{eq:limit_agg}) to (\ref{eq:Dr})
		\State Calculate $\mathcal{L}_s$ according to Equation ~(\ref{eq:social_score}) to (\ref{eq:Ds})
		\State Calculate $\mathcal{L} \leftarrow  \mathcal{L}_r + \lambda \mathcal{L}_s$
		\State Update parameters of $\mathcal{F}$ by using RSGD to optimize $\mathcal{L}$
		\Until{$\mathcal{L}$ converges or is sufficiently small}
		\State \textbf{return} {$\mathcal{F}(u, v | \Theta, \textbf{Y}, \textbf{S})$}
	\end{algorithmic}
	\label{alg:1}
\end{algorithm} 

\section{Experiments}  
\label{sec:experiment}

\subsection{Experiment Setup}
In this subsection, we introduce the datasets, baselines, evaluation protocols, and the choice of hyper-parameters.

\subsubsection{Datasets} 
We experimented with four datasets: Ciao\footnote{Ciao: http: //www.cse.msu.edu/\textasciitilde tangjili/index.html}, Yelp\footnote{Yelp: http://www.yelp.com/}, Epinion\footnote{Epinion: http://alchemy.cs.washington.edu/data/epinions/}, and Douban\footnote{Douban: http://book.douban.com}. Each dataset contains users’ ratings to the items and the social connections between users. In the data preprocessing step, we transform the ratings into implicit feedback (denoted by “1”) indicating that the user has rated the item positively. Then, for each user, we sample the same amount of negative samples (denoted by “0”) as their positive samples from unwatched items. Also, we filtered out users with no links in social networks. The statistics of the datasets are summarized in Table~\ref{exp:table1}.
\begin{table}[!htbp]
\begin{spacing}{0.9}
\centering  
\caption{Statistical details of the four datasets. “interactions” means user-item historical records, and “relations” denotes user-friend connections in social network.}
\begin{tabular}{{c|c c c c}}
\hline
dataset  & \# users & \# items & \# interactions & \# relations  \\ \hline\hline
Ciao     &7,210  &11,211 &147,590   &111,781 \\ \hline
Yelp     &10,580 &13,870 &342,204   &158,590 \\ \hline 
Epinion  &19,600 &23,585 &443,640   &351,485 \\ \hline
Douban   &12,748 &22,347 &1,570,544 &169,150 \\ \hline
\end{tabular}
\label{exp:table1}
\end{spacing}
\end{table}

\begin{table*}[!htbp]
\begin{spacing}{0.9}
	\centering
	\caption{The results of \textit{AUC} and \textit{Accuracy} in CTR prediction on four datasets. ** denotes the best values among all methods, and * denotes the best values among all competitors.}
	\begin{tabular}{ c | c c | c c | c c | c c }
		\hline
		\multirow{2}{*}{Method}&\multicolumn{2}{c|}{Ciao}&\multicolumn{2}{c|}{Yelp}&\multicolumn{2}{c|}{Epinion}&\multicolumn{2}{c}{Douban}\cr\cline{2-3}\cline{4-5}\cline{6-7}\cline{8-9}
		&AUC&ACC&AUC&ACC&AUC&ACC&AUC&ACC \cr
		\hline\hline
		FM       & 0.7538 & 0.6828 & 0.8153 & 0.7574 & 0.8048 & 0.7378 & 0.8304 & 0.7595 \\
		DeepFM   & \textbf{\ 0.7706*} & 0.6894 & 0.8302 & 0.7640 & 0.8158 & 0.7380 & 0.8347 & 0.7580 \\
		SoReg    & 0.7346 & 0.6663 & 0.8194 & 0.7546 & 0.7968 & 0.7269 & 0.8262 & 0.7481 \\
		TrustSVD & 0.7539 & 0.6765 & 0.8263 & 0.7567 & 0.8072 & 0.7404 & 0.8288 & 0.7607 \\
		DeepSoR  & 0.7646 & 0.6901 & 0.8308 & 0.7645 & 0.8171 & 0.7394 & 0.8303 & 0.7555 \\
		DiffNet  & 0.7704 & \textbf{\ 0.6924*} & \textbf{\ 0.8385*} & \textbf{\ 0.7686*} & 0.8162 & 0.7426 & 0.8295 & 0.7581 \\
		HOSR     & 0.7682 & 0.6895 & 0.8348 & 0.7615 & \textbf{\ 0.8219*} & \textbf{\ 0.7445*} & \textbf{\ 0.8353*} & \textbf{\ 0.7629*} \\ 
		\hline
		HSR      & \textbf{\ \ 0.7889**} & \textbf{\ \ 0.7183**} & \textbf{\ \ 0.8552**} & \textbf{\ \ 0.7919**} & \textbf{\ \ 0.8295**} & \textbf{\ \ 0.7580**} & \textbf{\ \ 0.8477**} & \textbf{\ \ 0.7743**} \\
		ESR    & 0.7645 & 0.6881 & 0.8330 & 0.7682 & 0.8063 & 0.7405 & 0.8307 & 0.7581 \\
		\hline
	\end{tabular}
	\label{exp:table2}
\end{spacing}
\end{table*}

\subsubsection{Comparison Methods}
To verify the performance of our proposed method HSR, we compared it with the following state-of-art social recommendation methods. The characteristics of the comparison methods are listed as follows:
\begin{itemize}[leftmargin= 13 pt]
\item \textbf{FM} is a feature enhanced factorization model. Here we utilize the social information as additional input features. Specifically, we concatenate the user embedding, item embedding and the average embeddings of user social neighbors as the inputs \cite{FM}.
\item \textbf{DeepFM} is also a feature enhanced factorization model, which combines factorization machines and deep neural networks. We provide the same inputs as FM for DeepFM \cite{DeepFM}.
\item \textbf{SoReg} models users' social relation as regularization terms to constrain the matrix factorization framework \cite{SocialReg}.
\item \textbf{TrustSVD} extends SVD++ \cite{SVD} framework by modelling both user-user social relations and user-item interaction \cite{TrustSVD}.
\item \textbf{DeepSoR} combines deep neural networks to learn users' preferences from social networks and integrate users' preferences into PMF \cite{PMF} framework for recommendation \cite{DeepSoR}.
\item \textbf{DiffNet} is a graph neural network based social recommender, which designs a layer-wise influence propagation structure for better user embedding modeling \cite{Diffnet}.
\item \textbf{HOSR} is a another graph neural network based social recommender, which propagates user embeddings along the social network and designs a attention mechanism to study the importance of different neighbor orders \cite{HOSR}.
\item\textbf{ESR} is the Euclidean counterpart of HSR, which replaces M{\"o}bius addition, M{\"o}bius matrix-vector multiplication, Gyrovector space distance with Euclidean addition, Euclidean matrix multiplication, Euclidean distance, and remove M{\"o}bius logarithmic map and exponential map.    
\item\textbf{HSR} is our complete model.
\end{itemize}

\subsubsection{Parameter Settings}
\par 

We implemented our methods with Pytorch which is a Python library for deep learning. For each dataset, we randomly split it into training, validation, and test sets following 7 : 1 : 2. The hyper-parameters were tuned on the validation set by a grid search. Specifically, the learning rate $\eta$ is tuned among $[10^{-4}, 5\times10^{-4}, 10^{-3}, 5\times10^{-3}]$; the embedding size $d$ is searched in $[8, 16, 32, 64, 128]$; the balancing factor $\lambda$ is chosen from $[10^{-4}, 10^{-3}, 10^{-2}, 10^{-1}]$; and the temperature $\tau$ is tuned amongst $[0.01, 0.05, 0.1, 0.5]$. For the aggregation layer size $L$, we set $L=1$ for all datasets and find them sufficiently good. We find using 2 or more layers can slightly increase the performance but take much longer training time. In addition, we set batch size $b=1024$, curvature $c=1$, social coefficient $\gamma=1$, and Fermi-Dirac decoder parameters $r = 2$, $t = 1$. We will further study the impact of key hyper-parameters in the following subsection. The best settings for hyper-parameters in all baselines are reached by either empirical study or following their original papers.

\subsubsection{Evaluation Protocols} 
We evaluate our methods in two scenarios: (i) in click-through rate (CTR) prediction, we adopt two metrics \textit{AUC} (area under the curve) and \textit{Accuracy}, which are widely utilized in binary classification problems; and (ii) in top-$K$ recommendation, we use the model obtained in CTR prediction to generate top-$K$ items. Since it is time-consuming to rank all items for each user in the evaluation procedure, to reduce the computational cost, following the strategy in \cite{NCF,Diffnet}, for each user, we randomly sample 500 unrated items at each time and combine them with the positive items in the ranking process. We use \textit{Precision@K} and \textit{Recall@K} to evaluate the recommended sets. We repeated each experiment 5 times and reported the average performance.

\subsection{Empirical Study}
\begin{figure}[h!]
	\centering
	\includegraphics[width=0.96\linewidth]{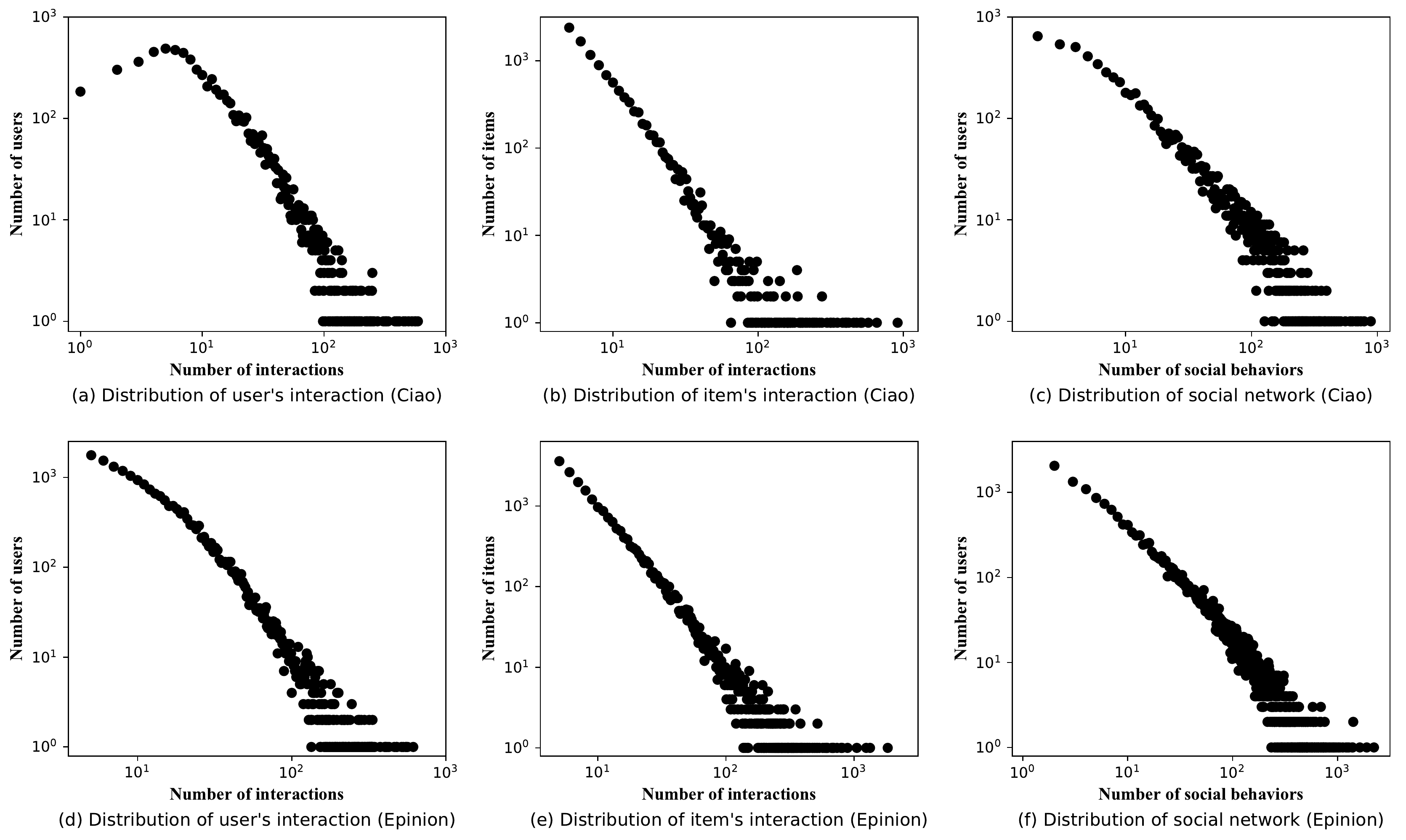}
	\caption{Distributions of user-item interaction and user-user social relations on Ciao (in top row) and Epinion datasets (in bottom row).}
	\label{fig:powlaw} 
\end{figure}

Researches show that data with a power-law structure can be naturally modeled in the hyperbolic space \cite{HHNE,Poincare,hp1}. Therefore, we conduct an empirical study to check whether the power-law distribution also exists in the user-item interaction and user-user social relations. For the user-item interaction relation, we present the distribution of the number of interactions for users and items, respectively. For the user-user social relation, we present the distribution of number of social behaviors for users. We show the distributions of these two relations on the Ciao and Epinion datasets in Figure~\ref{fig:powlaw}. We observed that these distributions show the power-law distribution: (i) for the user-item interaction relation, a majority of users/items have very few interactions and a few users/items have a huge number of interactions; and (ii) for the user-user social relation, a majority of users have very few social behaviors and a few users have a huge number of social behaviors. The above findings empirically demonstrate user-item interaction and user-user social relations exhibit power-law structure, thus we believe that using hyperbolic geometry might be suitable for the social recommendation.

\par

\begin{figure*}[h!]
	\centering
	\includegraphics[width=0.98\linewidth]{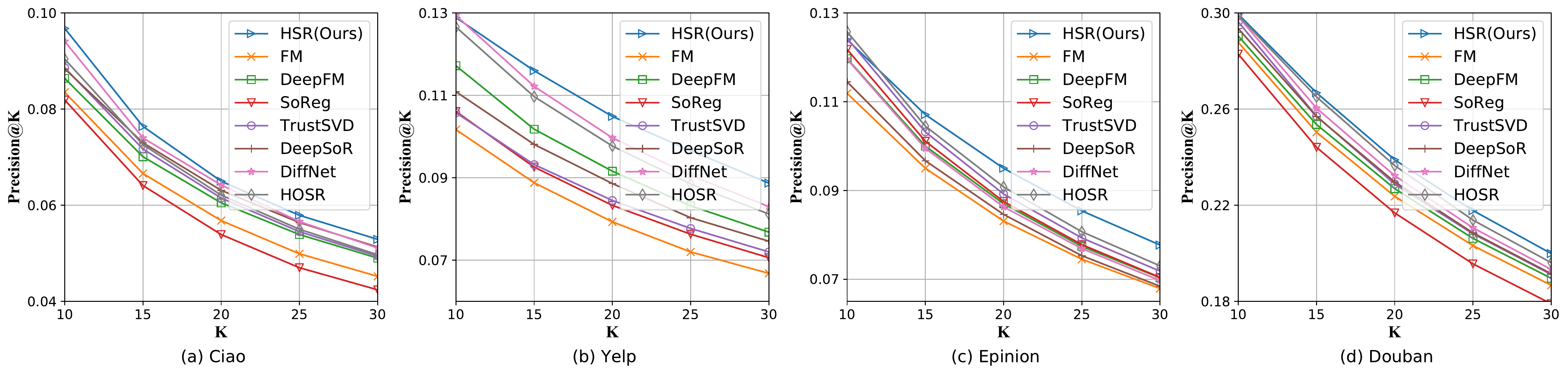}
	\caption{The results of \textit{Precision@K} in top-$K$ recommendation on four datasets}
	\label{fig:pre}
\end{figure*}

\begin{figure*}[h!]
	\centering
	\includegraphics[width=0.98\linewidth]{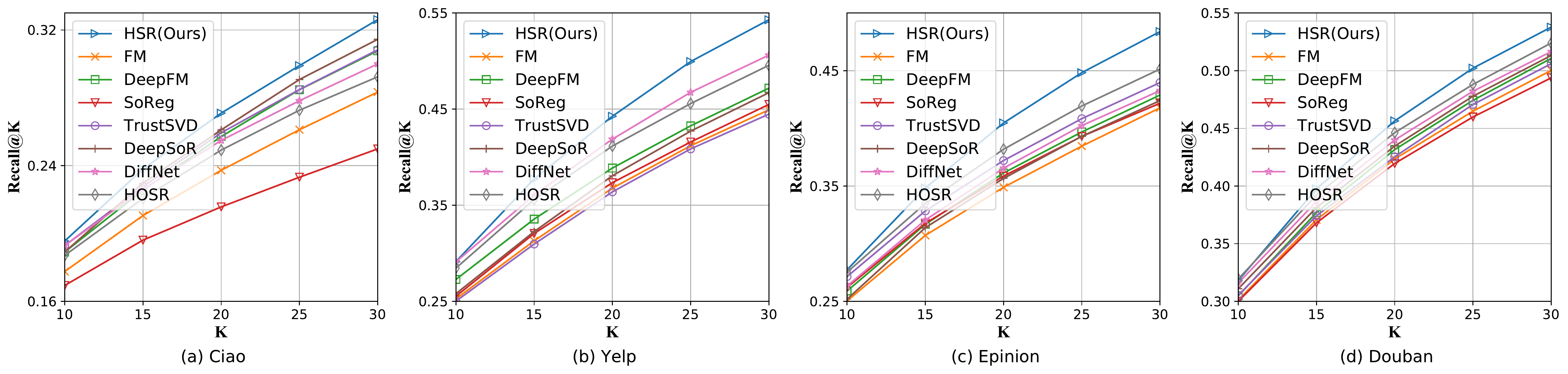}
	\caption{The results of \textit{Recall@K} in top-$K$ recommendation on four datasets}
	\label{fig:rec}
\end{figure*}

\subsection{Performance Comparison}
Table~\ref{exp:table2} and Figures~\ref{fig:pre}, ~\ref{fig:rec} show the performance of all compared methods in CTR prediction and top-$K$ recommendation, respectively (ESR are not plotted in Figure~\ref{fig:pre}, ~\ref{fig:rec} for clarity). From the results, we have the following main observations:
\par
(i) Deep learning based models (i.e., DeepFM, DeepSoR, DiffNet and HOSR) generally outperform shallow representation methods (i.e., FM, SoReg and TrustSVD), which indicates the effectiveness of applying neural components for social recommendation. 

(ii) Among baselines, graph neural network based social recommenders DiffNet and HOSR achieve strongly competitive performance. Such improvement verifies graph neural networks are powerful in representation learning for graph data, since it integrates the users' social information as well as users's topological structure in social network.

(iii) Intuitively, HSR has made great improvements over state-of-the-art baselines in both recommendation scenarios. For CTR prediction task, our method HSR consistently yields the best performance on four datasets. For example, HSR improves over the strongest baselines \textit{w.r.t.} \textit{Accuracy} by 3.73\%, 3.03\%, 1.81\%,and 1.49\% in Ciao, Yelp, Epinion and Douban datasets, respectively. In top-$K$ recommendation, HSR achieves 3.79\%, 5.68\%, 5.97\%, and 2.46\% performance improvement against the strongest baseline \textit{w.r.t.} \textit{Recall@20} in Ciao, Yelp, Epinion and Douban datasets, respectively. Considering the Euclidean counterpart of HSR, HSR achieves better scores than ESR. This indicates the effectiveness of social recommendation in the hyperbolic space.

\subsection{Handling Data Sparsity Issue}
\begin{figure}[h!]
	\centering
	\includegraphics[width=0.98\linewidth]{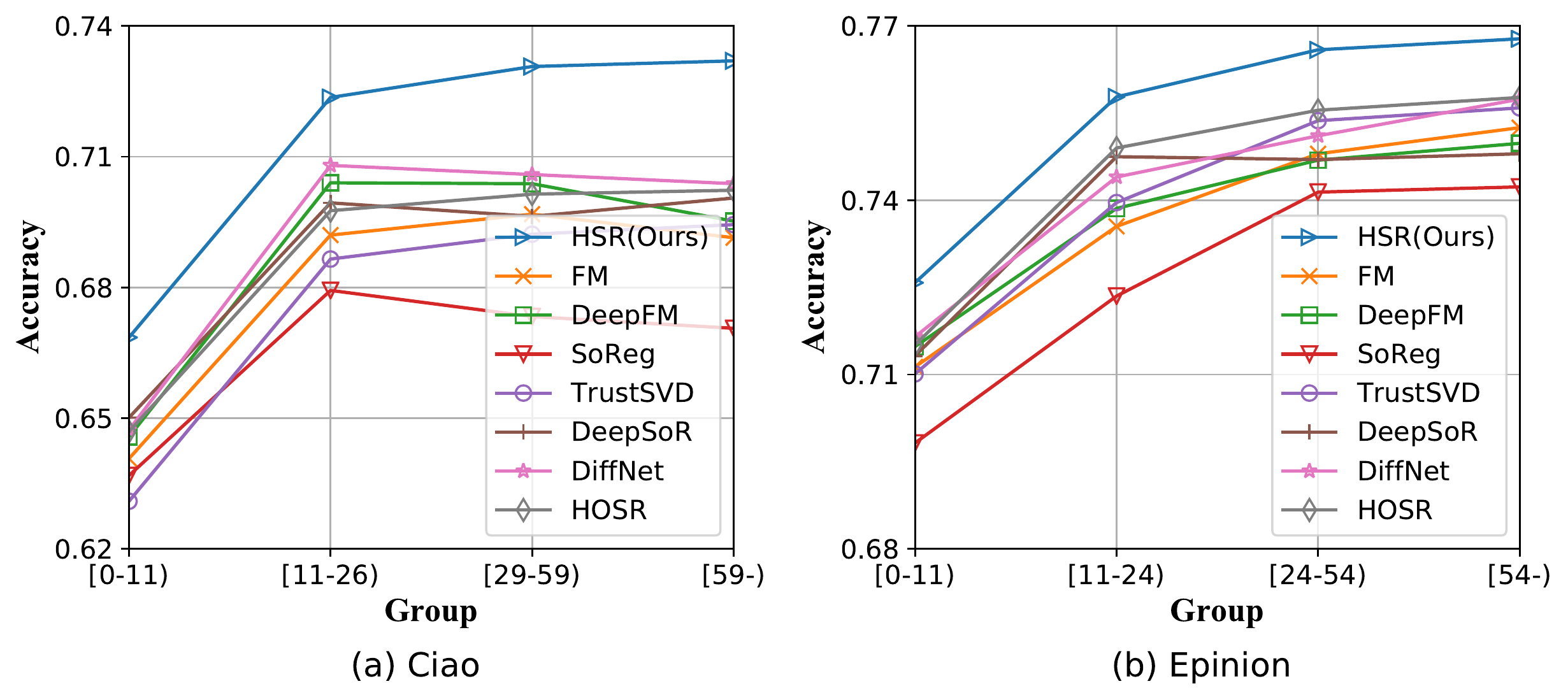}  
	\caption{Performance comparison over the sparsity distribution of user groups on Ciao and Epinion datasets.}
	\label{fig:sparse}
\end{figure}

The data sparsity problem is a great challenge for most recommender systems. To investigate the effect of data sparsity, we bin the test users into four groups with different sparsity levels based on the number of observed ratings in the training data, meanwhile keep each group including a similar number of interactions. For example, [11,26) in the Ciao dataset means for each user in this group has at least 11 interaction records and less than 26 interaction records. Figure~\ref{fig:sparse} shows the \textit{Accuracy} results on different user groups with different models on Ciao and Epinion datasets. From the results, we observe that HSR consistently outperforms the other methods including the state-of-the-art social enhanced methods like DiffNet and HOSR, which verifies that our method is able to maintain a decent performance in different sparse scenarios.

\subsection{Parameter Sensitivity}
\begin{figure}[h!]
	\centering
	\includegraphics[width=0.98\linewidth]{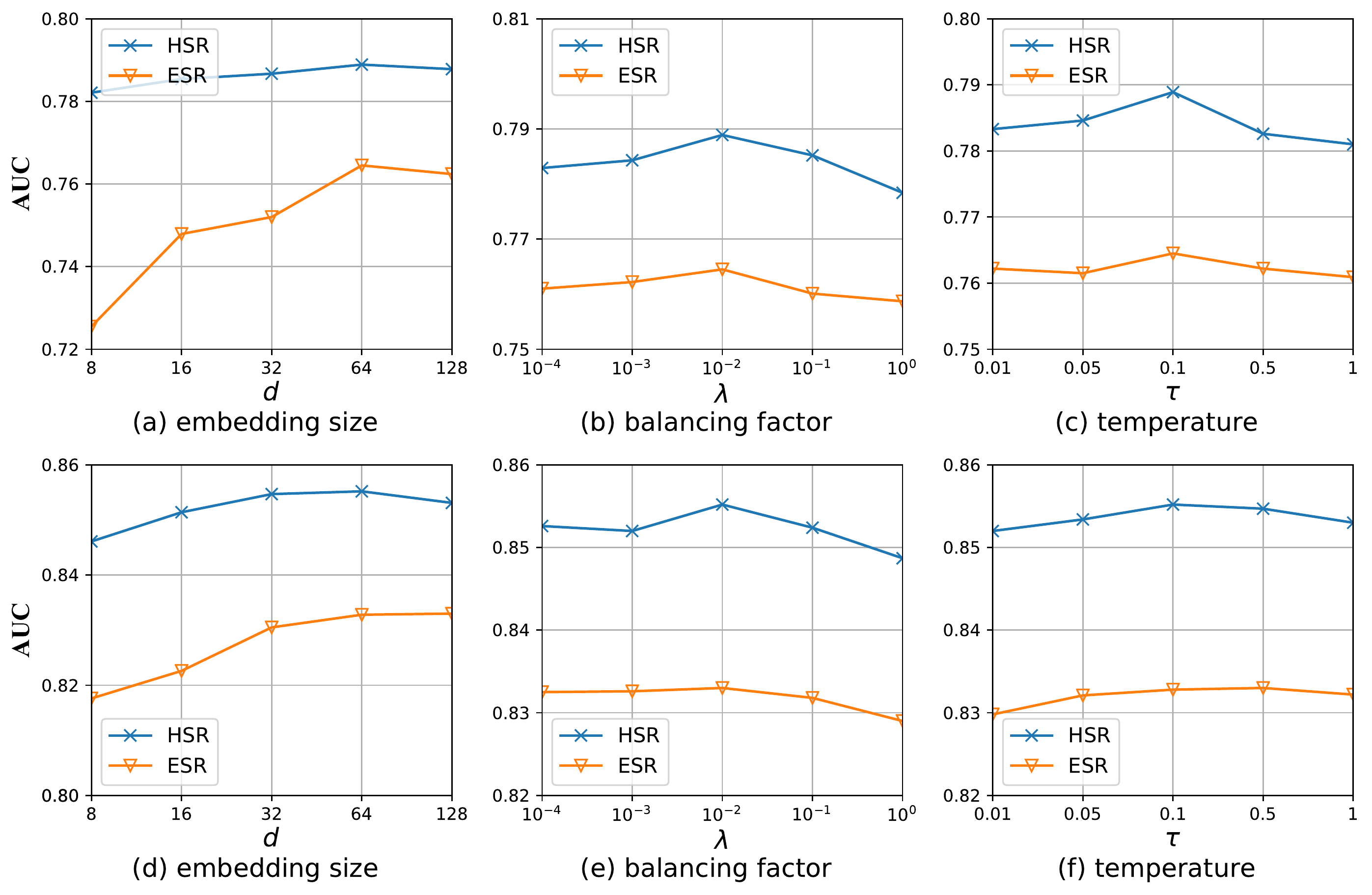}
	\caption{Parameter sensitivity on Ciao and Yelp datasets.}
	\label{fig:parameters}
\end{figure}

We explore the impact of four hyper-parameters: embedding size $d$, balancing factor $\lambda$, and temperature $\tau$. The results on Ciao and Yelp datasets are plotted in Figure~\ref{fig:parameters}. We have the following observations: (i) A proper embedding size $d$ is needed. If it is too small, the model lacks expressiveness, while a too large $d$ increases the complexity of the recommendation framework and may overfit the datasets. In addition, we observe that HSR always significantly outperforms ESR regardless of the embedding size, especially in a low-dimensional space, which shows that utilizing hyperbolic geometry can effectively learn high-quality representations for social recommendation. (ii) For balancing factor $\lambda$, we find that when $\lambda=10^{-2}$ is good enough on Ciao and Yelp datasets because a larger $\lambda$ will let the model focus on modeling social relation task. (iii) We find the empirically optimal temperature $\tau$ to be $0.1$ on Ciao and Yelp datasets. The performance increases when the temperature is tuned from 0 to the optimal value and then drops down afterwards, which indicates that proper value of $\tau$ can distill useful information for social neighbor aggregation. 

\subsection{Ablation Study}
To explore the effect of our neighbor attention mechanism and social relation modeling task, we conducted experiments with the two variants of HSR and ESR: (1) HSR-A and ESR-A (performing a mean operation in Equation ~(\ref{eq:eff_agg}), and (2) HSR-S and ESR-S (only modeling the recommendation task in Equation ~(\ref{equ:loss_high})). Table~\ref{exp:ablation} shows the \textit{AUC} results on four datasets. From the results, we find that removing any components will decrease recommendation performance of our models. For example, HSR-A performs worse than the complete model HSR, which shows that considering different influences of social neighbors in aggregation operation is necessary to model user preference. HSR also achieves better scores than HSR-S. This validates that explicitly modeling social relations is helpful for improving the model performance.

\begin{table}[h!]
\begin{spacing}{0.9}
\centering
\caption{Effect of the attention mechanism and social relation modeling on four datasets.}
\begin{tabular}{ c | c | c | c | c }  %
	\hline 
	Dataset  & Ciao   & Yelp   & Epinion & Douban  \\  \hline
	HSR      & 0.7889 & 0.8552 & 0.8295  & 0.8477  \\  
	HSR-A    & 0.7773 & 0.8490 & 0.8234  & 0.8431  \\  
	HSR-S    & 0.7784 & 0.8465 & 0.8247  & 0.8426  \\  \hline 
	ESR      & 0.7645 & 0.8330 & 0.8063  & 0.8307  \\  
	ESR-A    & 0.7593 & 0.8278 & 0.8002  & 0.8267  \\  
	ESR-S    & 0.7581 & 0.8295 & 0.8036  & 0.8253  \\  \hline 
\end{tabular}
\label{exp:ablation}
\end{spacing}
\end{table}

\subsection{Case Study}

\subsubsection{Embedding Analysis}

\begin{figure}[h!]
	\centering
	\includegraphics[width=0.98\linewidth]{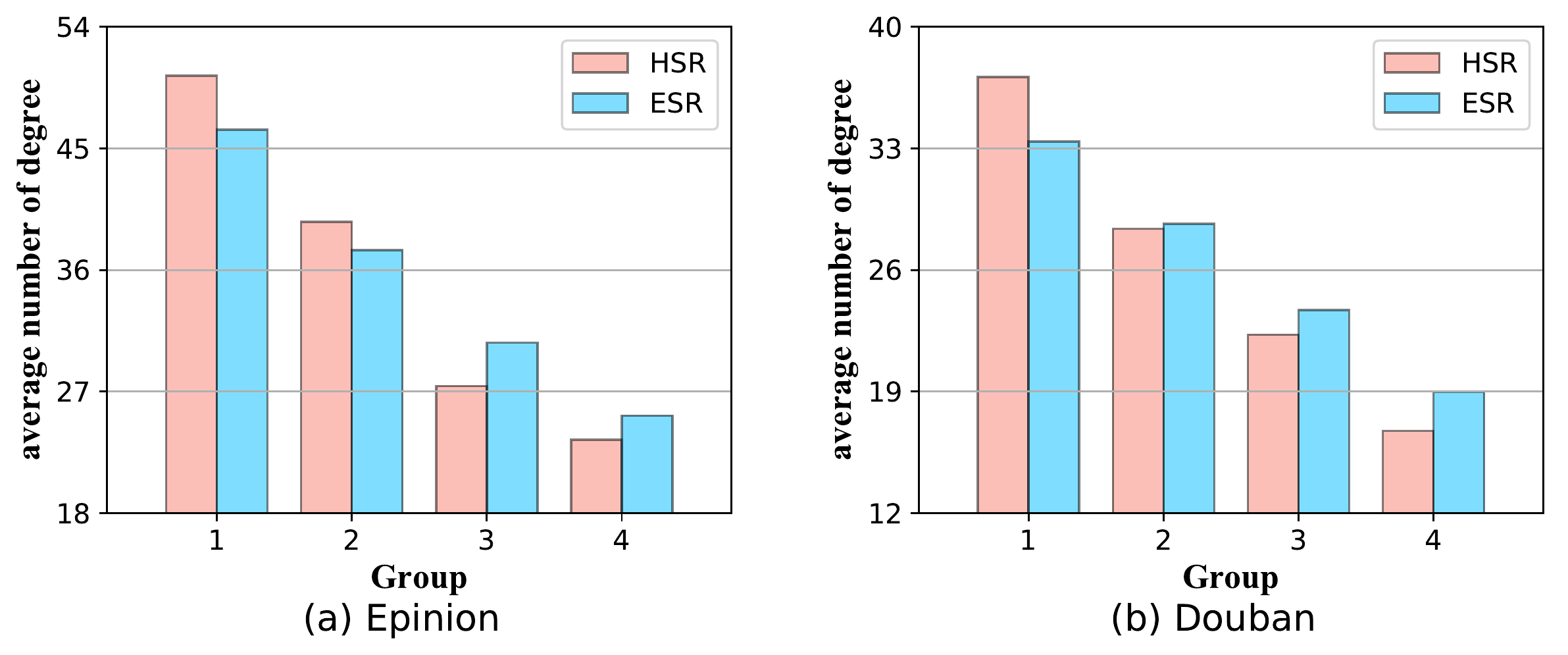}  
	\caption{Analysis of hierarchical structure in social networks on Epinion and Douban datasets.}
	\label{fig:region}
\end{figure}

Social networks often present a hierarchical structure \cite{social_h3,social_h2,social_h1}. In this case study, we evaluate whether our models can reflect hierarchical structures of social behaviors. In general, the distances between embeddings and the origin can reflect the latent hierarchy of graphs \cite{Poincare,HHNE}. We utilize Gyrovector space distance and Euclidean distance to calculate the distance to the origin for HSR and ESR, respectively. We bin the nodes in the user-user social graph into four groups according to their distances to the origin (from the near to the distant), meanwhile, keep each group including a similar number of nodes. For example, nodes in group 1 have the nearest distances to the origin while nodes in group 4 have the furthest distances to the origin. To evaluate the nodes' activity in the social graph, we compute the average number of nodes' social behaviors in each group. Figure~\ref{fig:region} shows the results on Epinion and Douban datasets. From the results, we can see that the average number of interaction behaviors decreases from group 1 to group 4. This result indicates that hierarchy of social behaviors can be modeled by our methods HSR and ESR. Compared with ESR, we find that HSR more clearly reflects the hierarchical structure, which indicates that hyperbolic space is more suitable than Euclidean space to embed data with the hierarchical structure.

\begin{figure}[h!]
	\centering
	\includegraphics[width=0.85\linewidth]{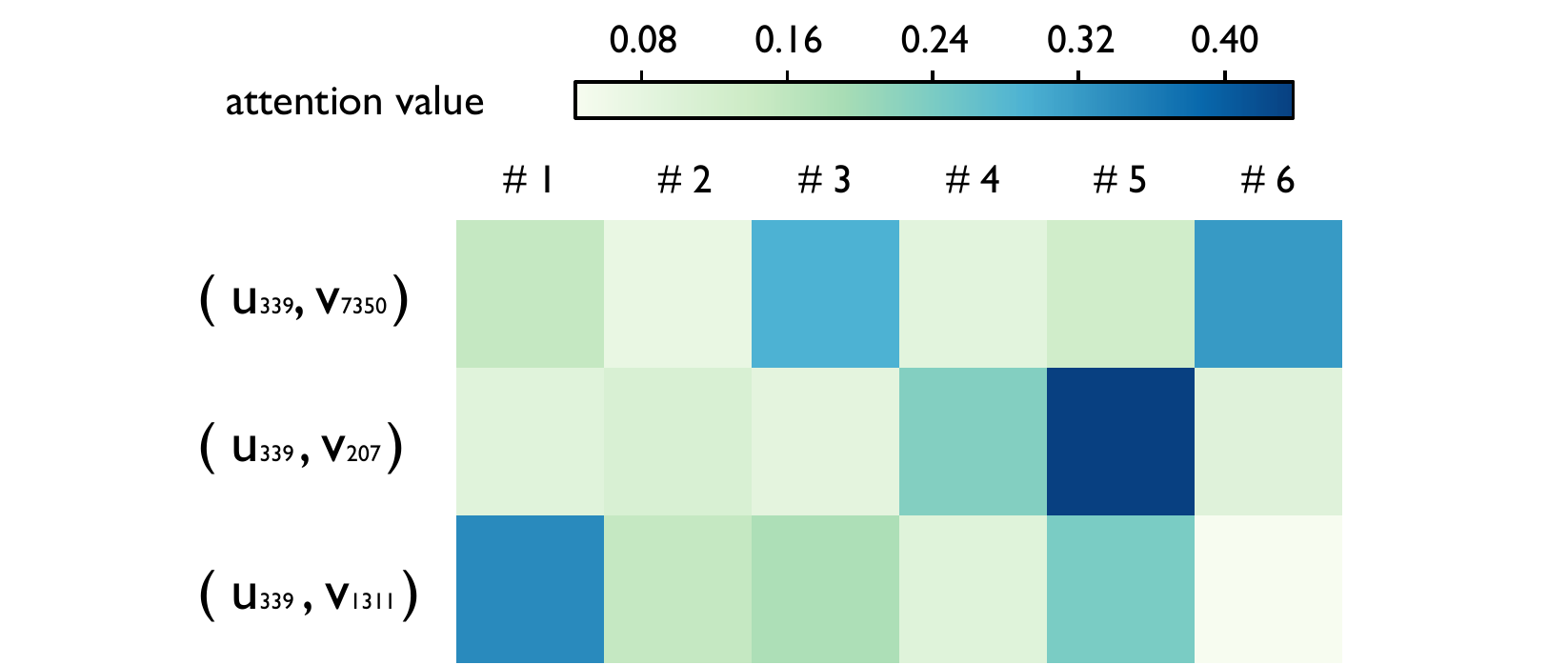}  
	\caption{Attention heatmap for the neighbors of three user-item pairs from Yelp dataset.}
	\label{fig:attention}
\end{figure}

\paragraph{Attention Analysis}
Benefiting from the attention mechanism, we can visualize the attention weights placed on the social neighbors for users, which reflects how the model learned. We randomly selected one user $u_{339}$ who has six friends from Yelp dataset, and three relevant items $v_{7350}$, $v_{207}$, $v_{1311}$ (from the test set). Figure~\ref{fig:attention} shows the attention weights of the user $u_{339}$'s social neighbors for the three user-item pairs. For convenience, we label neighbor ID starting from 1, which may not necessarily reflect the true ID from the dataset. From the heatmap, we have the following findings: (i) Not all neighbors have the same contribution when generating recommendations. For instance, for the user-item pair ($u_{339}$, $v_{7350}$), the attention weights of user $u_{339}$'s neighbor \# 3 and \# 6 are relatively high. The reason may be that neighbor \# 3 and \# 6 have rated item $v_{7350}$ in the training set. Therefore, neighbor \# 3 and \# 6 will provide more useful information when making recommendations. (ii) For different items, the attention distributions of the neighbors are different, which reflects the attention mechanism that can adaptively measure the influence strength of neighbors.

\section{Related Work}
\label{sec:related}
In this section, we provide a brief overview of two areas that are highly relevant to our work. 

\paragraph{Social Recommendation}

With the prevalence of online social media, social relations have been widely studied and exploited to boost the recommendation performance. The most common approach in social recommendation is to design loss terms for social influence and integrate both social loss and recommendation loss into a unified loss function for jointly optimizing \cite{SocialReg,CSR,SoDimRec}. For instance, SoReg assumes that social neighbors share similar feature representations and devise regularization terms to smooth connected users's embeddings \cite{SocialReg}; CSR models the characteristics of social relations and designs a characterized regularization term to improve SoReg \cite{CSR}. In addition to the regularization-based social recommenders, a more explicit and straightforward way is to explicitly models social relations in the predictive model \cite{RSTE,SPF,TrustSVD}. For example, TrustSVD extended SVD++ \cite{SVD} by incorporating each user’s social neighbors' preferences as the auxiliary feedbacks \cite{TrustSVD}. Inspired by the immense success of deep learning, some recent proposed social recommenders utilize neural components to enhance recommendation performance \cite{DeepSoR,SAMN,EATNN,Diffnet,HOSR}. For instance, DeepSoR combines a multi-layer perceptron to learn latent preferences of users from social networks with probabilistic matrix factorization \cite{DeepSoR}; SAMN considers both aspect-level and friend-level differences and utilizes memory network and attention mechanisms for social recommendation \cite{SAMN}; DiffNet stacks more graph convolutional layers and propagate users embeddings along the social network to capture high-order neighbor information \cite{Diffnet}. Different from above-mentioned social recommenders based on Euclidean representation methods, we propose a hyperbolic representation model HSR for social recommendation, which utilizes hyperbolic geometry to learn high-quality user and item representations in the non-Euclidean space.

\paragraph{Hyperbolic Representation Learning}
In recent years, representation learning in hyperbolic spaces has attracted an increasing amount of attention. Specifically, \cite{Poincare} embedded hierarchical data into the Poincar$\acute{\text{e}}$ ball, showing that hyperbolic embeddings can outperform Euclidean embeddings in terms of both representation capacity and generalization ability. \cite{hyperrelated1} focused on learning embeddings in the Lorentz model and showed that the Lorentz model of hyperbolic geometry leads to substantially improved embeddings. \cite{hyperrelated2} extended Poincar$\acute{\text{e}}$ embeddings to directed acyclic graphs by utilizing hyperbolic entailment cones. \cite{hyperrelated3} analyzed representation trade-offs for hyperbolic embeddings and developed and proposed a novel combinatorial algorithm for embedding learning in hyperbolic space. Besides, researchers began to combining hyperbolic embedding with deep learning. \cite{HNN} introduced hyperbolic neural networks which defined core neural network operations in hyperbolic space, such as M{\"o}bius addition, M{\"o}bius scalar multiplication, exponential and logarithmic maps. After that, hyperbolic analogues of other algorithms have been proposed, such as Poincar$\acute{\text{e}}$ Glove \cite{PoincareGlove} and hyperbolic attention networks \cite{HAT}.
\par
There are some recent works using hyperbolic representation learning for recommendation \cite{HyperBPR,HyperML,HME,HmusicRec,HEDoder,SHRS}. For instance, HyperBPR learns user and item hyperbolic representations and utilizes BPR \cite{BPR} framework for collaborative filtering \cite{HyperBPR}. HME is designed to solve next-Point-of-Interest (POI) recommendation task, which projects the check-in data into a hyperbolic space for representation learning \cite{HME}. Different from the above literature, we propose HSR which utilizes hyperbolic geometry for social recommendation. To our knowledge, HSR is the first work that uses the hyperbolic space for social recommendation tasks.

\section{Conclusion and Future Work}
\label{sec:conclusion}
In this work, we develop a novel method called HSR which leverages hyperbolic geometry for social recommendation. The key component of HSR is that we design a hyperbolic aggregator on the users’ social neighbor sets to take full advantage of the social information. We further introduce acceleration strategy and attention mechanism to improve our method HSR. Additionally, we also devise the social relation modeling task to make the user embeddings be reflective of the relational structure in social network, and jointly train it with recommendation task. To the best of our knowledge, HSR is the first model to explore the hyperbolic space in social recommendation. We conduct extensive experiments on four real-world datasets. The results demonstrate (i) the superiority of HSR compared to strong baselines, and (ii) the effectiveness of social recommendation in hyperbolic geometry. 

For future work, we will (i) extend our model for temporal social recommendation to consider users' dynamic preferences, and (ii) try to generate recommendation explanations for comprehending the user behaviors and item attributes. 

\newpage

\bibliographystyle{ACM-Reference-Format}
\bibliography{HSR}

\appendix

\end{document}